\documentstyle[11pt,newpasp,twoside,psfig]{article}
\begin{document}

\title{The Supranova Model and Its Implications}
\author{Arieh K\"onigl}
\affil{Department of Astronomy \& Astrophysics, University of
Chicago, 5640 S. Ellis Ave., Chicago, IL 60637, U.S.A.}

\setcounter{page}{1}
\index{Koenigl, A.}

\begin{abstract}
The supranova model for gamma-ray bursts, originally proposed by
Vietri \& Stella (1998), has
several unique features that make it attractive for the
interpretation of GRBs and their afterglows, and it has emerged 
as a promising candidate for modeling the evolution and  
environment of at least some GRB sources. This review summarizes
the model and its key observational implications, assesses its
strengths and potential weaknesses, and outlines paths for
future observational and theoretical work.
\end{abstract}

\section{Introduction}
There is mounting evidence that the progenitors of long-duration ($\ga 2\, {\rm s}$)
gamma-ray bursts (GRBs) are massive stars that collapse to form stellar-mass
black holes or strongly magnetized and rapidly rotating neutron
stars. Up until recently, the prevalent
scenario has been the ``collapsar'' model, in which the GRB event
directly follows the stellar collapse (e.g., Woosley, these
Proceedings). This picture is consistent with the identification in a few
GRBs of a possible signature of a nearly simultaneous supernova Ib/Ic explosion.
In one commonly discussed realization of this scenario, the
duration of the burst is determined by the accretion
time of the unexpelled stellar material through the debris disk
that forms around the newly created central black hole (BH).

Vietri \& Stella (1998) proposed an alternative to the collapsar
picture, dubbed the ``supranova'' model, in which the supernova
(SN) explosion initially results in the formation of a comparatively
massive, magnetized neutron star (NS) endowed with rapid
rotation. This ``supramassive'' NS (SMNS) is
envisioned to gradually lose rotational support through a
pulsar-type wind\footnote{It is assumed in what follows that angular
momentum loss by gravitational waves remains relatively unimportant.}
until it eventually becomes unstable to gravitational collapse
(leading to the formation of a BH and the triggering of a GRB). The
original motivation for the supranova model was the desire to
identify a comparatively ``baryon clean'' environment
in which the high ($\ga 100$) Lorentz factors inferred in the
prompt high-energy emission region could be attained. (In the
proposed picture, the gas around the GRB progenitor is swept out by
the expanding SN ejecta over the spindown time of the SMNS.)
As discussed below, subsequent considerations of this
scenario have revealed that it also has the potential of
providing a natural interpretation of several key observational
features of GRB afterglows. 
\section{The Supranova Scenario}
\subsection{SMNS Evolution}

An SMNS is a general-relativistic
equilibrium configuration of a uniformly rotating NS
whose mass exceeds (by up to $\sim 20\%$, for realistic equations
of state) the maximum mass of a nonrotating NS (e.g., Cook et
al. 1994; Salgado et al. 1994).\footnote{Differential
rotation may support a significantly larger mass against gravitational
collapse, allowing the formation of a {\em hypermassive} NS
(e.g., Baumgarte et al. 2000). However, a differentially rotating NS is
estimated to evolve to a uniform rotation profile (through
magnetic field amplification and viscosity) on a timescale much shorter than
the spindown time of a uniformly rotating star (e.g., Shapiro
2000). Therefore, in contrast with
supramassive configurations, hypermassive neutron stars above
the supramassive mass limit are not expected to exhibit significant delays between
NS formation and collapse to a BH.} An SMNS that
loses angular momentum and energy adiabatically while conserving
its total baryon mass follows an evolutionary sequence that
brings it to a point where it becomes unstable to axisymmetric
(and possibly first to nonaxisymmetric) perturbations, which
result in collapse to a BH. 

For an SMNS of mass $M_*$, radius $R_*$, angular velocity
$\Omega_*$, disposable (before the SMNS collapses) rotational
energy $\Delta E_{\rm rot} = \alpha G M_*^2
\Omega_*/2 c \approx 10^{53}\ {\rm ergs}$, and surface magnetic
field $B_*$, the spindown time due to a pulsar-type wind is
\begin{displaymath}
t_{\rm sd} = 4 \left
( {\alpha \over 0.5}\right ) \left ({M_* \over 2 \,
M_\odot}\right )^2
\left ( {R_* \over 15 \ {\rm km}} \right )^{-6} \left ({\Omega_*
\over 10^4\ {\rm s}^{-1}}\right )^{-3} \left ({B_* \over 10^{12}\
{\rm G}}\right )^{-2}\ {\rm yr}\ ,
\end{displaymath}
which can range from several weeks to several years for
plausible values of $M_*$, $R_*$, and $B_*$ ($\sim 10^{12}-10^{13}\ {\rm G}$).

One possible scenario of how the formation of a BH could
trigger the GRB outburst is that a transient debris disk
comprising the outer layers of the SMNS is left behind when the
bulk of the SMNS collapses, and that the stellar magnetic field
threading this material is amplified by differential rotation to $\ga
10^{14}\ {\rm G}$ and drives a Poynting flux-dominated outflow
from the disk surfaces (e.g., M\'esz\'aros \& Rees 1997;
Vlahakis \& K\"onigl, these Proceedings).

\subsection{Theoretical Foundations of the Model}

Only a small fraction of newly formed neutron stars could belong
to the supramassive class: this is consistent with the fact that
GRBs are rare events, but so far our knowledge of how a rotating
star collapses is not sufficient to make an {\em a priori}
determination of this fraction. However, the results already available
suggest that this fraction may well be nonnegligible. For example,
numerical simulations of {\em nonrotating} stars have been
used to infer that the distribution of NS masses
produced through core collapse may be flat beyond
the $1.2-1.6\, M_\odot$ range where it peaks (Fryer \& Kalogera
2001). Not unexpectedly, when rotation is included, the simulations yield more
massive remnants (Fryer \& Heger 2000). The NS angular velocity
distribution derived from stellar-evolution calculations is sensitive to
the adopted assumptions. These studies indicate that, in the absence of magnetic
angular-momentum transport, newly formed neutron stars will
likely rotate near breakup (Heger et al. 2000). Recently proposed prescriptions
for the inclusion of magnetic effects (e.g., Spruit 2002)
imply strong angular-momentum transport and comparatively low initial values
of $\Omega_*$ (Heger \& Woosley 2002), but their validity
remains uncertain. In fact, scaling arguments from white-dwarf
systems suggest that core-to-envelope angular momentum transport
in pulsar progenitors may be rather inefficient  (Livio \& Pringle
1998).

Another relevant question is whether a debris disk --- perhaps
the most likely origin of a GRB outflow -- is left behind when
the SMNS eventually collapses to a BH. Although a collapsing supramassive compact object
with a soft equation of state could leave
behind up to $\sim 10\%$ of its mass in the form of a
rotationally supported disk (e.g., Shapiro \&
Shibata 2002), realistic SMNS equations of state might be too
stiff in this regard, with the implication that the entire SMNS
would collapse (e.g., Shibata et al. 2000). However,
comparatively stiff SMNS configurations could reach the
mass-shedding limit before they become dynamically unstable (e.g., Cook et
al. 1994), so in principle such stars might also give rise to a
BH surrounded by a highly magnetized debris
disk.\footnote{Although additional material (e.g., some of the
fallback from the original SN explosion) could surround the
newly formed BH, this gas would not be as highly magnetized as
the SMNS matter and therefore might not contribute significantly
to the GRB outflow.}

\section{GRB Afterglows in Pulsar-Wind Bubbles}

The supranova scenario also has important implications to GRB
afterglows. In particular, the bubble created by the SMNS pulsar-type wind provides an
environment that can account for both the high fraction ($\epsilon_e$) of the
internal energy residing in relativistic electrons and positrons
and the high magnetic-to-internal energy ratio ($\epsilon_B$) that
have been inferred in the afterglow emission regions of a number
of sources through spectral modeling. The deduced
values of $\epsilon_e$ ($\ga 0.1$; e.g., Panaitescu \& Kumar
2002) are hard to explain in a standard [interstellar-medium (ISM) or
stellar-wind] environment, especially for long-lasting
afterglows. The values of $\epsilon_B$ are typically inferred
to be not much smaller than $\epsilon_e$  (for example, spectral
fits to the GRB 970508 afterglow have yielded $\epsilon_B$
estimates in the range $\sim 0.01-1$). These estimates are again
problematic for GRB outflows that propagate into a standard
ambient medium, since $\epsilon_B$ will only be $\sim 10^{-10}$
for a Milky Way-like ISM and at most $\sim 10^{-4}$ for a magnetized
stellar-wind environment.

The comparatively high values of $\epsilon_e$ and $\epsilon_B$
can, however, arise naturally in a pulsar-wind bubble (PWB),
since the composition of relativistic pulsar
winds is likely dominated by an $e^+e^-$ component, and since
such winds are often characterized by a high magnetization parameter  $\sigma_w =
B_w^2/4\pi\rho_wc^2$ (the Poynting-to-particle energy flux
ratio, where $B_w$ is the magnetic field amplitude and $\rho_w$ is the
rest-mass density, both measured in the wind frame). This has
been explicitly demonstrated by K\"onigl \& Granot (2002;
hereafter KG02).

KG02 constructed a PWB model based on a number of
simplifications ---  including spherical symmetry, negligible
ambient mass, a steady state, pure $e^+e^-$
composition, a monoenergetic pair distribution at each radius, and
a dominant azimuthal magnetic-field component. In their representation,
the SMNS-driven pulsar-type wind shocks
at a radius $R_s$ and fills the volume between $R_s$ and
the radius $R_b$ of the swept-up SN-ejecta shell with a relativistic
fluid  consisting of ``hot'' $e^+e^-$ pairs and electromagnetic
fields. The ejecta shell (of mass $M_{\rm ej} \approx 10 M_\odot$)
is accelerated by the bubble (becoming compressed and fragmented
in the process) and expands at a speed $v_b \sim
(\Delta E_{\rm rot}/M_{\rm ej})^{1/2}\approx 0.1c$. The afterglow emission is
attributed to the bow shock of a GRB jet that propagates between
$R_s$ and $R_b$. KG02's choice of wind
parameters was guided by recent observations of the Crab and
Vela SN remnants (SNRs), and their model incorporates synchrotron-radiation
cooling (parameterized by $a_1 \approx$ synchrotron cooling
time in units of $R_b/c$)\footnote{As was pointed out by Guetta \&
Granot (2003; hereafter GG03), the effect of cooling on the PWB
structure is less important if ions dominate the wind particle
energy flux.} and a possible departure from ideal MHD (associated
with an ``equipartition'' upper bound on the
electromagnetic-to-thermal pressure ratio) in the shocked-wind zone.

KG02 showed that their model can account for the inferred
values of $\epsilon_e$ and $\epsilon_B$ as well as for the deduced range
of ambient densities in GRB afterglows (e.g., Panaitescu \&
Kumar 2002). In this interpretation, the ambient density is not
dominated by the baryon rest mass as in the
standard picture, but rather by the relativistic
inertia of the shocked wind and compressed electromagnetic fields, so the effective
preshock hydrogen number density is given by
\begin{displaymath}
n_{\rm H,equiv} =\frac{4p+(B
+E)^2/4\pi}{m_{\rm p}c^2}\, ,
\end{displaymath}
where $p$, $B$, and $E$ are, respectively, the particle pressure and the magnetic
and electric field amplitudes in the PWB (all measured in the fluid rest frame).
By using the parameterization $n_{\rm
H,equiv}(r) \propto r^{-k}$, it was found that $k$ takes on a range
of values (determined by $\sigma_w$, $a_1$, and the location
within the bubble) that encompass both a uniform-ISM and a
stellar-wind radial density profiles. The derived solutions have
demonstrated that in the context of this scenario it is possible to explain
how a GRB with a massive stellar progenitor can produce an
afterglow that shows no evidence of either a stellar wind or a
high-density environment, which has been hard to understand in the
standard picture. Additional observational implications of the PWB
interpretation are discussed by Guetta \& Granot in these Proceedings.

\section{Interpretation of X-ray Features}

As reviewed in one of this Workshop's sessions, several
afterglow sources have shown evidence for X-ray emission (and absorption)
features, which have been identified as either Fe or primarily
``light''-element (Mg, Si, S, Ar, Ca)
transitions. Although it has been recognized early on 
that these features are consistent with an origin in a large-scale SNR
shell and could thus lend support to the supranova model,
alternative explanations that might be compatible with the
collapsar model have also been proposed (e.g., Kallman et
al. 2001). In the following paragraphs I first summarize the
interpretation of the reported emission features
that was outlined in KG02, and then I discuss its implications
to the supranova scenario.

In the supranova picture, the emission is induced by continuum
irradiation from the central
region around the time of the burst. Light-travel effects limit
the solid angle from within which emission is received at time
$t$ to $\Delta\Omega(t)/{4\pi} = [1-\cos{\theta(t)}]/2 =
ct/2(1+z)R_b$, where $z$ is the source redshift. For example, in
the case of GRB 991216 (observed between 37 and $40.4\, {\rm
hr}$; Piro et al. 2000), the model implies that
the detected emission originated at polar angles $\theta \sim
27^{\circ}-28^{\circ}$ with respect to the jet axis. The
emission feature is interpreted as Fe XXV He$\alpha$ from a photoionized gas of
ionization parameter $\xi\equiv L_i/nR_b^2\approx 10^3$, where
$L_i$ is the ionizing continuum luminosity. For $L_i=4\pi D^2 F_x =
6.1\times10^{45}\ {\rm ergs\ s^{-1}}$, $M_{\rm Fe}\approx 0.1\, M_\odot$, and
$T\approx 10^6\, {\rm K}$, one infers
$R_b\approx 2\times 10^{16}\, {\rm cm}$. Since the expansion
time to $R_b \approx 10^{16}\, {\rm cm}$ at a speed
$v_b \approx 0.1\, c$ is $\sim 10^2$ days, this estimate is consistent with
the bulk of the SN-ejected radioactive $^{56}$Ni decaying into
$^{56}$Fe. The estimated density is $n \approx 2\times 10^{10}\, {\rm
cm^{-3}}$, which likely corresponds to clumps in the
Rayleigh-Taylor unstable SNR shell. If the clumps have a
covering factor $\sim 1$ and an iron abundance a few times
solar, then the implied Thomson optical depth $\tau_T$ is $\la 1$ and the Fe
photoionization optical depth is a few, which are optimal for producing
high--equivalent-width iron lines through
reflection.\footnote{Reflection from a
medium with $\tau_T\gg 1$ is, however, also a
possibility; e.g., Ballantyne et al. (2002).} 

The identification of X-ray features in a source like GRB 011211
with ``light'' elements (Reeves et al. 2002) can be interpreted in this picture in terms
of a bubble characterized by $t_{\rm sd} \sim {\rm
weeks}$ (rather than months), in which iron has not
yet formed in the shell (Granot \& Guetta 2002).

The value of $R_b$ inferred from modeling the X-ray emission feature in
GRB 991216 is smaller (by about an order of magnitude) than the distance
deduced from spectral fits to the continuum afterglow
emission. As was pointed out by KG02, a plausible
resolution of this apparent discrepancy is that, rather than
being spherical, the PWB is, in fact, {\em elongated} along the
SMNS rotation axis. (Further discussion of this
possibility, as well as additional arguments against a simple
spherical model, are given in GG03.) Such a morphology is a likely
outcome of the PWB structure and of its origin. In particular,

\noindent
${\bullet}$ The supernova explosion of a rapidly
rotating progenitor star is likely to be nonspherical, and so is
also the pulsar wind;

\noindent
${\bullet}$ Both the SN ejecta and the environment into which the
SNR expands (which was possibly shaped by earlier episodes of stellar mass
loss during the red-supergiant and blue-supergiant evolutionary
phases of the progenitor star) could have a highly anisotropic mass distribution;

\noindent
${\bullet}$ Magnetic hoop stresses may act to
strongly collimate the PWB (and possibly also the preceding
stellar outflows), an effect enhanced by cooling (e.g.,
Chevalier \& Luo 1994; Gardiner \& Frank 2001).

In this interpretation, a narrow GRB jet moves within the PWB
along the rotation axis, and the afterglow shock reaches much larger
distances from the center than the spherical portion of the SNR
shell from which the X-ray emission features originate. This
picture could potentially be corroborated by
observations.\footnote{Given, however, that a variety of factors could in
principle prevent the effects discussed here from being observed, a
failure to detect them in any given source would not necessarily
constitute evidence against this interpretation.} For example, it implies (on account of
light-travel effects) a minimum onset time for
the detectability of X-ray emission features. There have already been
several reported null detections of such features in
sensitive X-ray observations of GRB sources: the above picture suggests that
continuing the search in sources like these at later times
might be worth while. Another implication of the above interpretation
is that the polar region of the bubble expands faster (by a
factor of up to $\sim 10$) than the lower-latitude SNR shell where the
emission features are produced. If a dense SN-ejecta clump
located near the major axis of the PWB was carried out
by the expanding bubble and happened to intercept our line of sight, then
an absorption feature with an implied blueshift significantly larger than that
of the emission feature might be detected.\footnote{A preliminary report
on a tentative detection of such an absorption feature in GRB
011211 was given in F. Frontera's presentation in this Workshop.}
\section{Assessment of the Model}

Several recent papers that scrutinized the supranova scenario
have drawn attention to a number of potential weaknesses of this model. I
now briefly summarize, and comment on, some of the issues that
have been raised.

\noindent
${\bullet}$ Occurrence of long bursts. It has been argued
(e.g., B\"ottcher \& Fryer 2001) that the supranova scenario is most
likely to produce short bursts, rather than the long bursts to
which this model has been applied. There are, however, potential
alternatives to ongoing mass feeding (as in the collapsar model)
for producing long disk lifetimes. They include a low disk
viscosity (e.g., Popham et al. 1999; Ruffert \& Janka 1999) and
a magnetically mediated spin-up torque
exerted by the BH (van Putten \& Ostriker 2001).

\noindent
${\bullet}$ Apparent evidence (from the detection of a reddened
bump in the GRB lightcurve in a few sources) for a SN/GRB
coincidence. If this interpretation is correct, it argues against a time delay
between the SN explosion and the GRB event. However,
alternative interpretations of these bumps have been proposed,
including illumination of ambient dust by the GRB source (e.g.,
Esin \& Blandford 2000; Waxman \& Draine 2000) and (in the
context of the supranova scenario) thermal emission from a
PWB-heated SNR shell (Dermer 2002). Furthermore, as discussed by
GG03, the SN/GRB time delay may span a range of values, with a
near coincidence (as in the collapsar model) possibly
characterizing a subset of sources.

\noindent
${\bullet}$ Disruption of SNR shell. In contrast to the PWB model
described in \S 3, Inoue et al. (2003) explored a scenario
wherein the pulsar wind disrupts and penetrates through the
SN-ejecta shell, producing a bubble whose size is about an order of
magnitude larger than the SNR radius. One can argue, however,
that since $v_b$ as calculated in the KG02 picture is much less
than the wind speed ($\sim c$), it is likely that the wind will
shock and remain confined within the SNR shell. Existing
numerical simulations of ``plerionic'' SNRs (e.g., Jun 1998) seem to be
consistent with this conclusion, but additional studies that
explicitly consider the case where the total wind energy greatly
exceeds that of the SN and that explore strong departures from
spherical symmetry would be useful.

\noindent
${\bullet}$ Apparent need for fine-tuning. McLaughlin et
al. (2002) observed that the typical shell size ($\sim 10^{16}\,
{\rm cm}$) inferred in the supranova interpretation of the detected X-ray
emission features translates
into a characteristic value of the SN/GRB time delay: they
questioned why this value would be constant from source to
source and wondered whether it would be long enough to account
for the decay of nickel in the SN ejecta into iron. However, as was
noted in \S 4, there is now evidence for X-ray features arising in ``light'' elements
rather than iron, which can be
naturally interpreted in the supranova picture in terms of lower
values of $t_{\rm sd}$. As was also noted
in \S 4, the time delays required for iron to
become dominant are consistent (in view of the $\sim
0.1c$ shell expansion speed estimated in the spherical model) with $R_s$ being $\ga
10^{16}\, {\rm cm}$. Furthermore, as was pointed out by KG02, shells with
sizes much larger than $10^{16}\, {\rm cm}$ would have densities
that are too low to give rise to observable features. These
considerations can
explain why the inferred radii of iron-emitting shells are
roughly the same. Another fine-tuning issue was raised by
Woosley et al (2002), who called attention to the fact that SMNS configurations
correspond to a limited NS mass range, especially in comparison
with hypermassive neutron stars (in which significant SN/GRB
time delays are not expected; see footnote 2). Although
this statement is factually correct, it seems to be more of a
description of a constraint that the model needs to satisfy
(which may be part of the reason why GRBs are rare events) than
of a serious flaw.

In summary, even though the ultimate verdict on this model will be provided
by additional observations and theoretical calculations, it
appears that none of the objections raised against it so far is compelling.

\section{Conclusions}
The supranova model shows great
promise in that it naturally addresses several key modeling
issues of GRBs and their afterglows:

\noindent
$\bullet$ a relatively baryon-free near-source
environment (the pulsar-wind cavity) in which a high-Lorentz-factor outflow can
be accelerated;

\noindent
$\bullet$ a pair-rich and potentially strongly
magnetized environment further out (the pulsar-wind bubble) in
which a high-$\epsilon_e$ and $\epsilon_B$ afterglow emission can arise;

\noindent
$\bullet$ a heavy element-enriched, extended dense
medium (the clumpy SNR shell) in which the X-ray spectral
features can be produced.

The model can readily account for a variety of observed behaviors in GRB afterglows, which may be
classified according to the magnitude of the SMNS spin-down time
$t_{\rm sd}$ (expected to vary in the range of weeks
to years). For example, X-ray features associated
with ``light'' elements might arise when $t_{\rm
sd}\sim {\rm weeks}$, whereas Fe features could be
found when $t_{\rm sd}\sim {\rm months}$.

Simple preliminary estimates
concerning the physical viability of this picture are
encouraging, but more detailed studies of the formation and
evolution of supramassive neutron stars and of the pulsar-wind
bubbles that they drive are needed. Several
unique observational predictions could serve to
discriminate between this model and alternative scenarios.

\end{document}